\documentclass[aps,prb,twocolumn,amsmath,superscriptaddress]{revtex4}
\usepackage{graphicx}
\usepackage{color}
\usepackage{epstopdf}
\usepackage{xcolor}

\begin{document}

\title{Charge fluctuations of a Cr atom probed in the optical spectra of a quantum dot}

\author{L. Besombes}\email{lucien.besombes@neel.cnrs.fr}
\affiliation{Univ. Grenoble Alpes, CNRS, Grenoble INP, Institut N\'{e}el, 38000 Grenoble, France}

\author{H. Boukari}
\affiliation{Univ. Grenoble Alpes, CNRS, Grenoble INP, Institut N\'{e}el, 38000 Grenoble, France}

\author{V. Tiwari}
\affiliation{Univ. Grenoble Alpes, CNRS, Grenoble INP, Institut N\'{e}el, 38000 Grenoble, France}

\author{A. Lafuente-Sampietro}
\affiliation{Univ. Grenoble Alpes, CNRS, Grenoble INP, Institut N\'{e}el, 38000 Grenoble, France} \affiliation{Institute of Materials Science,
University of Tsukuba, 1-1-1 Tennoudai, Tsukuba 305-8573, Japan}

\author{M. Sunaga}
\affiliation{Institute of Materials Science, University of Tsukuba, 1-1-1 Tennoudai, Tsukuba 305-8573, Japan}

\author{K. Makita}
\affiliation{Institute of Materials Science, University of Tsukuba, 1-1-1 Tennoudai, Tsukuba 305-8573, Japan}

\author{S. Kuroda}
\affiliation{Institute of Materials Science, University of Tsukuba, 1-1-1 Tennoudai, Tsukuba 305-8573, Japan}

\date{\today}

\begin{abstract}

We study the emission of individual quantum dots in CdTe/ZnTe samples doped with a low concentration of Cr. In addition to dots with a
photoluminescence (PL) split by the exchange interaction with a magnetic Cr atom, we observe another type of dots with a complex PL structure
composed of a minimum of six lines on the exciton and biexciton and three lines on the charged excitons. In these dots, the linear polarization
dependence and the magnetic field dependence of the PL behave like three similar quantum dots emitting at slightly different energies.
Cross-correlation intensity measurements show that these emission lines are not independent but exchange intensities in a time scale of a few
hundred nanoseconds depending on the optical excitation power. We attribute this PL structure to charge fluctuations of a Cr atom located in the
vicinity the CdTe dots in the ZnTe barrier. We present a model which confirms that the presence of a single charge fluctuating between -e
(Cr$^{+}$), 0 (Cr$^{2+}$) and +e (Cr$^{3+}$) and located a few nm away from the dot explains the observation of three emission energies. We finally
show that the interaction between the confined carriers and the nearby fluctuating localized charge can be modified by an applied static electric
field which modulates the splitting of the emission lines.

\end{abstract}

\maketitle

\section{Introduction}

Individual spins in semiconductors are a promising platform for the development of emerging quantum technologies with solid state devices
\cite{Koenrad2011}. The activity in this field is mainly focused on the spin of individual carriers in quantum dots \cite{Veldhorst2015} (QD) or
strongly localized spins on individual defects \cite{Schmitt2017}. Individual magnetic atoms in diluted magnetic semiconductor offering a localized
electronic spin with a significant exchange interaction with the carriers of the host are also of interest. The spin state of individual or pairs of
magnetic atoms inserted in a QD can be controlled optically \cite{Besombes2004,Kudelski2007,Goryca2009,LeGall2010,LeGall2011,Besombes2012,Krebs2013}
thanks to their exchange interaction with the confined carriers. A large variety of magnetic atoms can be incorporated in semiconductors giving a
large choice of electronic spin, nuclear spin as well as orbital momentum \cite{Kobak2014,Smolenski2016}.

Magnetic atoms with an orbital momentum present a large spin to strain coupling and could find application as spin $qubits$ in spin/nano-mechanical
systems \cite{Barfuss2015,Lemonde2018}. This is the case of Cr which carries an electronic spin S=2 and an orbital momentum L=2 \cite{Vallin1974}
when it is incorporated in substitution in II-VI semiconductors as a Cr$^{2+}$ ion. In this neutral state, the bonding orbitals are filled and the
inner d shell has an open configuration d$^4$. Under optical excitation or in the presence of electrical dopants, charge transfer to the band of the
semiconductor or to other localized levels may however occur leading to changes in the d shell configuration and in the charge state of the Cr. If
the transition metal capture an electron, the impurity acquires a charge -e with respect to the lattice and the configuration and oxidation state
become d$^5$ and Cr$^{1+}$ respectively. Symmetrically, after capture of a hole, the charge state is +e and the configuration and oxidation states
are d$^3$ and Cr$^{3+}$. Depending on the host semiconductor, the associated acceptor or donor levels can appear in the band gap \cite{Rzepka1993}.

Magnetic CdTe/ZnTe QDs containing an individual Cr$^{2+}$ ion (Cr atom in the d$^4$ configuration) were recently studied
\cite{Lafuente2016,Lafuente2017Cr}. The photoluminescence (PL) of neutral Cr-doped QDs is dominated by three main excitonic emission lines
corresponding to the spin states $S_z$=0 and $S_z$=$\pm$1 of the Cr exchange-coupled with the spins of the confined electron and hole. The central
line, associated with S$_z$=0, is usually split by the electron-hole exchange interaction and linearly polarized along two orthogonal directions.
This optical emission structure together with its magnetic field dependence results from the exchange interaction between the spin of the confined
carriers and the spin of the magnetic atom and permits to clearly identify CdTe dots containing a Cr atom \cite{Lafuente2018}.

We show in this article that in Cr-doped CdTe/ZnTe QD samples, another type of QDs presenting an emission on the neutral or charged excitons
consisting of complex multiplets cannot be explained by the exchange interaction of the confined carriers with the spin of a Cr atom. Linear
polarization analysis and magneto-optics studies show that these QDs behave like three similar dots emitting at slightly different energies. We
attribute this PL structure to charge fluctuations of a Cr atom located in the ZnTe barrier near a given CdTe dot. The electric field produced by the
three possible charge states of the Cr in ZnTe (Cr$^{+}$, Cr$^{2+}$, and Cr$^{3+}$)\cite{Kuroda2007} results in three different shifts of the
confined carriers energies through the Stark effect.

This article is organized as follows: after a short presentation of the samples and experiments in Sec. II, we describe in Sec. III the PL structure
of individual QDs submitted to the charge fluctuations of a Cr atom and compare them with the emission of dots containing a Cr$^{2+}$ ion. We discuss
in Sec. IV their linear polarization and magnetic field dependence. In Sec. V, we present cross and auto-correlation intensity measurements
confirming that the lines observed in these emission spectra arise from the same QD. In Sec VI and VII we show how the electric field produced by a
single Cr atom located in the ZnTe barriers can induce a detectable energy shift of the emission of single QD. Finally, in Sec VIII we demonstrate
that the influence of the charge fluctuations of a Cr atom on a nearby QD can be modified by the optical excitation intensity, the optical excitation
wavelength and an applied uniform static electric field.

\section{Samples and experiments}

The studied sample consists of Cr-doped self-assembled CdTe QDs grown by molecular beam epitaxy on a p-doped ZnTe (001) substrate
\cite{Wojnar2011,Lafuente2016}. A low density of Cr is chosen to avoid multiple Cr to interact with the carriers confined in a given QD. A bias
voltage can be applied between a semi-transparent 5 nm gold Schottky gate deposited on the surface of the sample and the p-doped ZnTe substrate to
create a static electric field across the Cr-doped QDs layer.

The PL of individual QDs is studied by optical micro-spectroscopy at low temperature (T=5K). The PL is excited with a continuous wave ($cw$) dye
laser tuned to an excited state of the QDs, dispersed and filtered by a 1 $m$ double spectrometer before being detected by a Si cooled multichannel
charged coupled device (CCD) camera. A magnetic field, up to 11 T, can be applied along the QD growth axis.

The temporal statistics of single dot emission is analyzed through photon-correlation measurements using a Hanbury Brown and Twiss (HBT) setup with a
time resolution of about 0.8 ns \cite{Sallen2010}. Under our experimental conditions, with photon count rates of a few kHz, the photon pair time
distribution given by the HBT measurements yields, after normalization, to the second order correlation function of the PL intensity,
$g^{(2)}(\tau)$. In all these experiments, a high refractive index hemispherical solid immersion lens is deposited on top of the sample to enhance
the collection of the single-dot emission.

\section{PL structure of individual QDs in Cr-doped CdTe/ZnTe samples.}

\begin{figure}[hbt]
\includegraphics[width=3.25in]{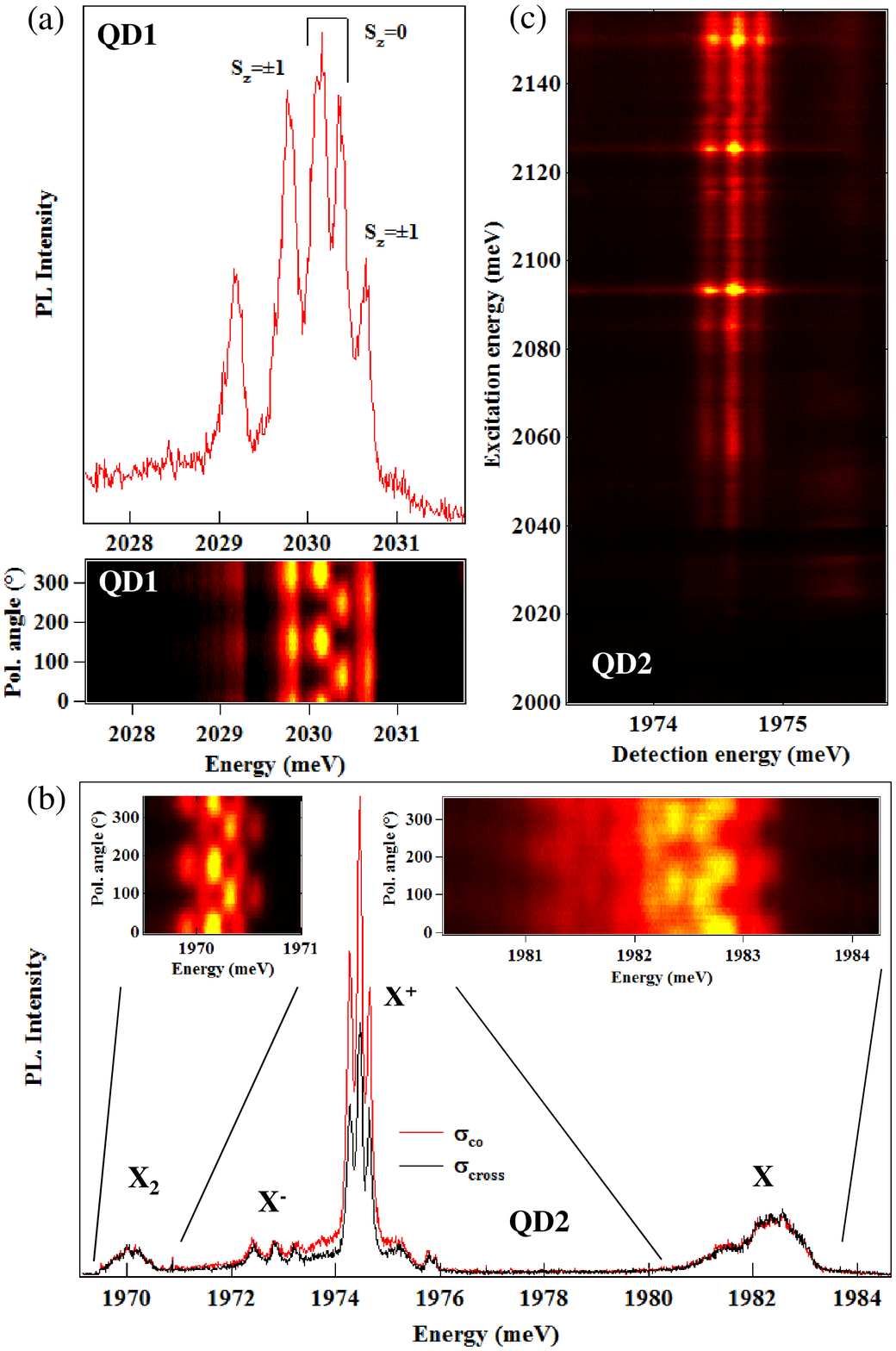}
\caption{Example of QDs (QD1 and QD2) illustrating the different PL structure observed in the studied Cr-doped sample. (a) PL of QD1 containing a Cr atom.
(b) Co and Cross circularly polarized PL of QD2 presenting a complex linear polarization structure on the exciton (biexciton) and three lines on the charged excitons.
 The insets show the linear polarization intensity maps of the exciton and biexciton. (c) PL excitation spectra recorded on the positively charged exciton of QD2.} \label{Fig1}
\end{figure}

Individual CdTe/ZnTe QDs containing a single Cr$^{2+}$ ion (S=2) where recently developed \cite{Lafuente2016,Lafuente2017Cr,Lafuente2018}. An example
of low temperature (T=5K) PL of such QD observed in the studied charge tunable sample is presented in Fig.~\ref{Fig1}(a) (QD1). It is dominated by
three main emission lines associated to the Cr spin states S$_z$=0 (pair of central lines) and S$_z$=$\pm1$ (two outer lines). The thermalization in
the presence of a large strain induced magnetic anisotropy prevents the observation of the highest energy spin states S$_z$=$\pm2$
\cite{Lafuente2016}. In this QD the central line is split and linearly polarized along two orthogonal directions (Fig.~\ref{Fig1}(a), bottom panel).
This structure arises from the broken cylindrical symmetry of the QD which is common and due to anisotropic in-plane strain and/or elongated shape.
It corresponds to the fine structure splitting induced by the electron-hole exchange interaction usually observed in non-magnetic QDs
\cite{Bayer2002}. The influence of the electron-hole exchange interaction is reduced on the two outer lines as the corresponding bright excitons are
already split by their exchange interaction with the Cr spin. The outer lines are then weakly linearly polarized. An additional PL line from a dark
exciton state also appears on the low energy side of the spectra. This PL structure, together with its magnetic field dependence \cite{Lafuente2018},
is a characteristic signature of the carrier-Cr spins exchange interaction and permits to clearly identify QDs containing a Cr$^{2+}$ ion with a
significant overlap with the confined carriers.

In the Cr-doped QD samples with a targeted Cr concentration two or three times larger than the one used previously to detect magnetic QDs containing
a single Cr \cite{Lafuente2018}, broad emission lines with a complex linear polarization structure are often observed in the micro-PL spectra for
excitons (X) and biexcitons (X$_2$). In most of these dots, the positively charged exciton (X$^+$) can also be observed at zero bias voltage. An
example, QD2, is presented in Fig. \ref{Fig1}(b). X$^+$ is easily identified by its position in energy, its absence of linearly polarized fine
structure and its large co-circularly polarized emission under resonant excitation on an excited state of the dot. It presents a clear structure of
three PL lines. As observed in the PLE spectra of QD2 (Fig.~\ref{Fig1}(c)), the three emission lines of X$^+$ have common excited states. As within a
sub-micron laser excitation spot the probability to find three independent charged dots with similar PL intensities, regular energy spacing and
common excited states is very unlikely, these lines unambiguously arise from the same QD. This will be confirmed by cross-correlation intensity
measurements. As illustrated by the different examples presented in this paper, similar spectra are observed in many dots.

\begin{figure}[hbt]
\includegraphics[width=3.25in]{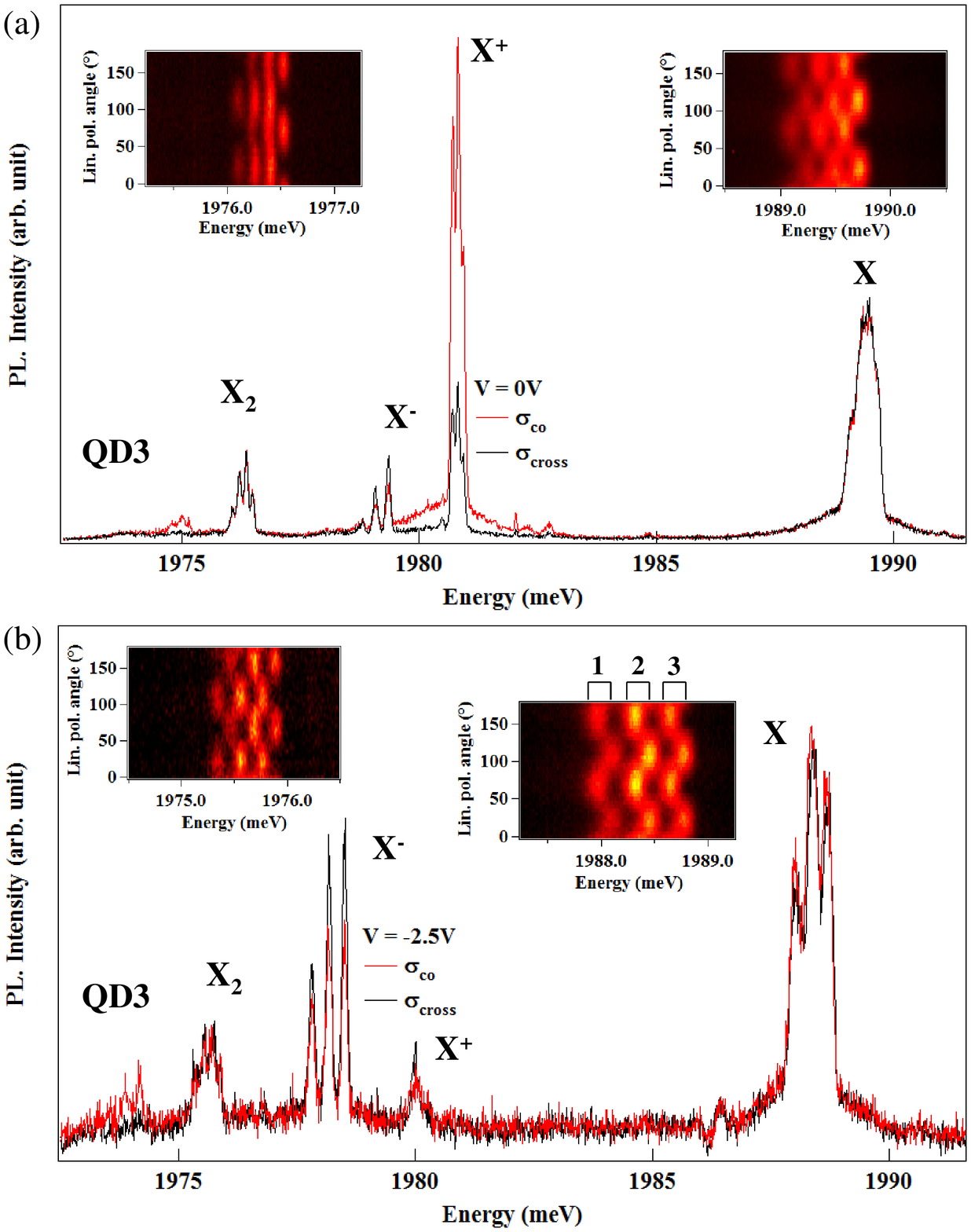}
\caption{Co and cross-circularly polarized PL of QD3 at zero applied electric field (a) and under an applied voltage V=-2.5V (b).
 The insets show the corresponding linear polarization intensity maps of the exciton and biexciton.
 The structure of three doublets on the neutral species is clearly revealed by the linear polarization analysis under an applied electric field.}
\label{Fig2}
\end{figure}

In some of these dots, a detailed analysis of the linear polarization dependence of the neutral species (X or X$_2$) shows that the spectra are
dominated by three doublets linearly polarized along two orthogonal directions. The linearly polarized spectra of such dot (QD3) is presented in
Fig.~\ref{Fig2}. At zero applied bias voltage, the emission of QD3 is dominated by the three lines of X$^+$ and the neutral species present broad
emission lines with some linear polarization dependence. Applying a weak bias voltage V=-2.5V  (Fig.~\ref{Fig2}(b)) decreases the contribution of the
X$^+$ and slightly increases the overall line-width of X and X$_2$. The PL of X and X$_2$ have a slightly different overall splitting. However, under
an electric field they clearly present a PL structure form of three doublets (labeled 1, 2 and 3 on the linear polarization intensity map of X in the
inset of Fig.~\ref{Fig2}(b)). The lines of each doublet are linearly polarized along two orthogonal directions. A mirror symmetry is observed in the
directions of linear polarization of each doublet for X and X$_2$. Such mirror symmetry in the linear polarization of X and X$_2$ is common in
non-magnetic QDs and is a consequence of the radiative cascade occurring during the radiative recombination of the biexciton \cite{Bacher1999}.

Let us note that in the micro-spectroscopy spectra dots of this type are most of the time identified thanks to the emission of X$^+$ composed of
three well resolved lines. For the neutral QDs, the splitting of the neutral exciton into six lines which are not well resolved usually leads to a
broad emission peak not easy to assigned to the type of QDs considered here (see X in QD2 in Fig.~\ref{Fig1}(b) and X in QD3 in Fig.~\ref{Fig2}(a)).

\begin{figure}[hbt]
\includegraphics[width=3.25in]{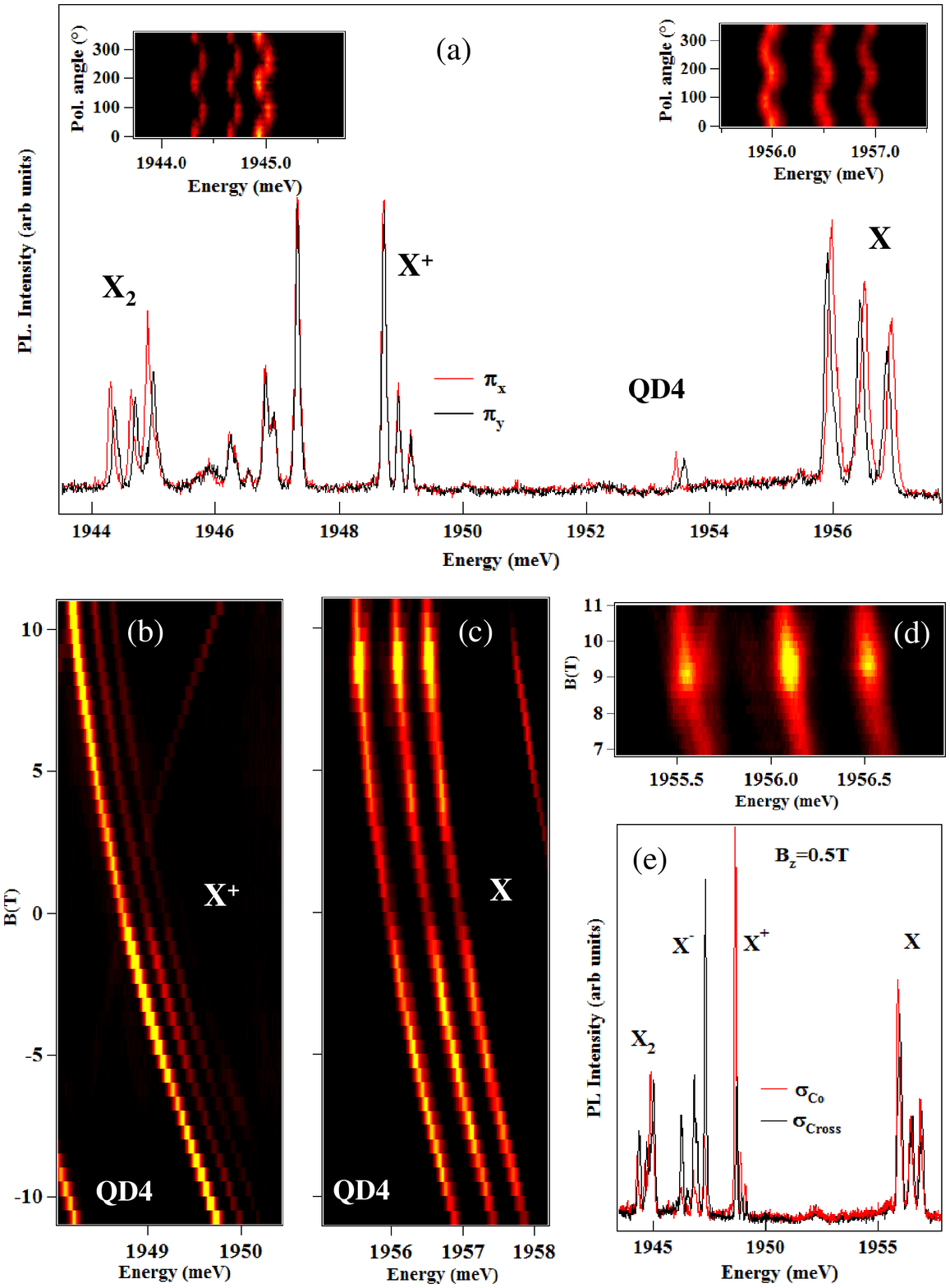}
\caption{(a) Linearly polarized PL spectra of QD4. The insets present the corresponding linearly polarized PL intensity map of X and X$_2$. Six lines (three doublets) are clearly observed on the neutral species.
(b) Magnetic field dependence of the PL of X$^+$. (c) Magnetic field dependence of the PL of X in QD4.
(d) Zoom on the anti-crossings observed in the magnetic field dependence of X around B$_z$=9T. (e) Co and cross-circularly polarized PL spectra under a magnetic field B$_z$=0.5T.
The large co-circularly (cross-circularly) polarized emission permits to identify the contributions of the positively (negatively) charged exciton.} \label{Fig3}
\end{figure}

This linear polarization dependence for the exciton and the biexciton (and the charged exciton which has no linear polarization) is similar to the
one that could be obtained for three similar QDs emitting at slightly different energies. The linear polarization structure would then be due to the
electron-hole exchange interaction in an anisotropic QD which splits the neutral species and does not affect the charged excitonic complexes
\cite{Bayer2002}.

\section{Magnetic field dependence of the PL structure of QDs in Cr-doped samples.}

Among this type of QDs, selecting a dot having a small splitting of the linearly polarized doublets of the exciton (for instance QD4 in
Fig.~\ref{Fig3}(a)) permits to clearly analyze the evolution of the PL under magnetic field. The magnetic field dependence of X$^+$ and X in such QD
is presented in Fig. \ref{Fig3}(b) and (c) respectively. The magnetic field dependence of X strongly differs from the case where a Cr atom is located
inside the dot (see reference \cite{Lafuente2018}). The intensity map presented in Fig.~\ref{Fig3}(c) does not show the characteristic anti-crossings
of Cr-doped QDs. For this type of QDs, each linearly polarized doublet of X has a similar evolution with a linear Zeeman energy shift which increases
the zero field exchange induced splitting. The emission becomes circularly polarized and an anti-crossing is observed in $\sigma-$ polarization
around B$_z$=9T on each line. Such anti-crossings under large magnetic field can be observed in standard non-magnetic QDs \cite{Leger2007}. They
arise from a mixing of bright and dark excitons which are shifted by the Zeeman energy and mixed by the electron-hole exchange interaction. Such
mixing can occur in low symmetry QDs (symmetry lower than C$_{2v}$: a truncated lens shape QD for instance). For the X$+$ which is not affected by
the electron-hole exchange interaction, a simple Zeeman splitting is observed under magnetic field for each of the three PL lines.

As already pointed out in the analysis of the linear polarization dependence at zero field, the observed magnetic field dependence of the exciton and
the charged exciton is similar to the one expected for three QDs having slightly different emission energies and a weak zero field fine structure
splitting.

\section{Fluctuations of the PL intensity of QDs observed in cross and auto-correlation measurements.}

To confirm that in this type of spectra the different lines arise from the same QD, we used the statistics of time arrivals of the emitted photons
given by the second order correlation function of the PL intensities g$^{(2)}(\tau)$ \cite{Sallen2010}.

\begin{figure}[hbt]
\includegraphics[width=3.25in]{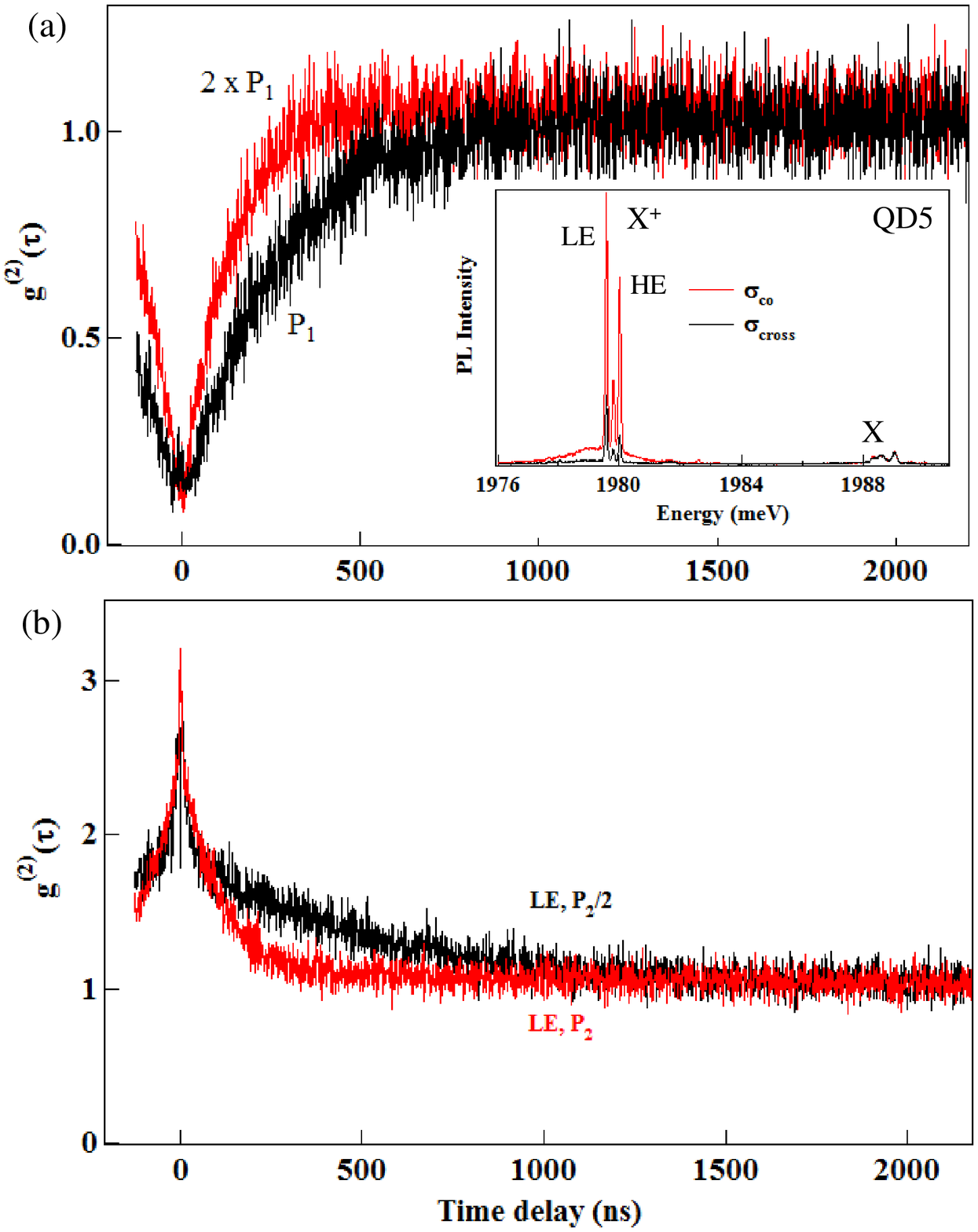}
\caption{(a) Cross-correlation of the PL intensity of the high energy (HE) and low energy (LE) lines of the positively charged exciton in QD5. The inset presents the co and cross-circularly polarized PL of QD5.
(b) Auto-correlation of the PL intensity of the high energy line of QD5 for different excitation intensities.}
\label{Fig4}
\end{figure}

Cross-correlation measurements using a HBT set-up were first performed on the circularly polarized PL emitted by the high energy (HE) and the low
energy (LE) lines of a positively charged exciton (QD5, see inset of Fig.~\ref{Fig4}(a)). In these experiments, the excitation is tuned to an excited
state of the QD at 2076 meV. The cross-correlation presented in Fig.~\ref{Fig4}(a) shows a clear anti-bunching with a FWHM in the 250 ns range and
g$^{(2)}$(0)$\approx$0.1. The presence of this anti-bunching confirms that the lines are not independent as they cannot simultaneously emit photons:
they arise from the same single photon emitter. The width of the cross-correlation signal corresponds to the transfer time between these two energy
levels of the same QD. As presented in Fig. \ref{Fig4}(a), this transfer which occurs in a hundred ns time-scale is accelerated by the increase of
the excitation power.

Similar fluctuations in the intensities of the lines of the charged exciton can be measured with auto-correlation measurements (Fig. \ref{Fig4}(b)).
Whereas the cross-correlation probes the transfer time between two energy levels, the auto-correlation give the time dependence of the probability
for the QD emission energy to be conserved. A large bunching with a width in the hundred ns is observed for instance in the auto-correlation in the
low energy line. This results from fluctuations of the PL intensity with a rate similar to the one observed in the cross-correlation measurements.
These fluctuations also significantly depend on the excitation power (Fig. \ref{Fig4}(b)) and should occur in a much longer time-scale in the absence
of optical excitation.

\section{Possible charges states of a Cr atom in ZnTe}

A possible origin for a three lines spectra in an individual QD is the presence of an electric field which fluctuates between three different
discrete values during the PL integration time. Such PL structure of three lines is only observed in CdTe/ZnTe QD samples doped with Cr. We propose
that this fluctuating electric field arises from the presence of a Cr atom in the ZnTe barrier close to the studied QD.

\begin{figure}[hbt]
\includegraphics[width=3.0in]{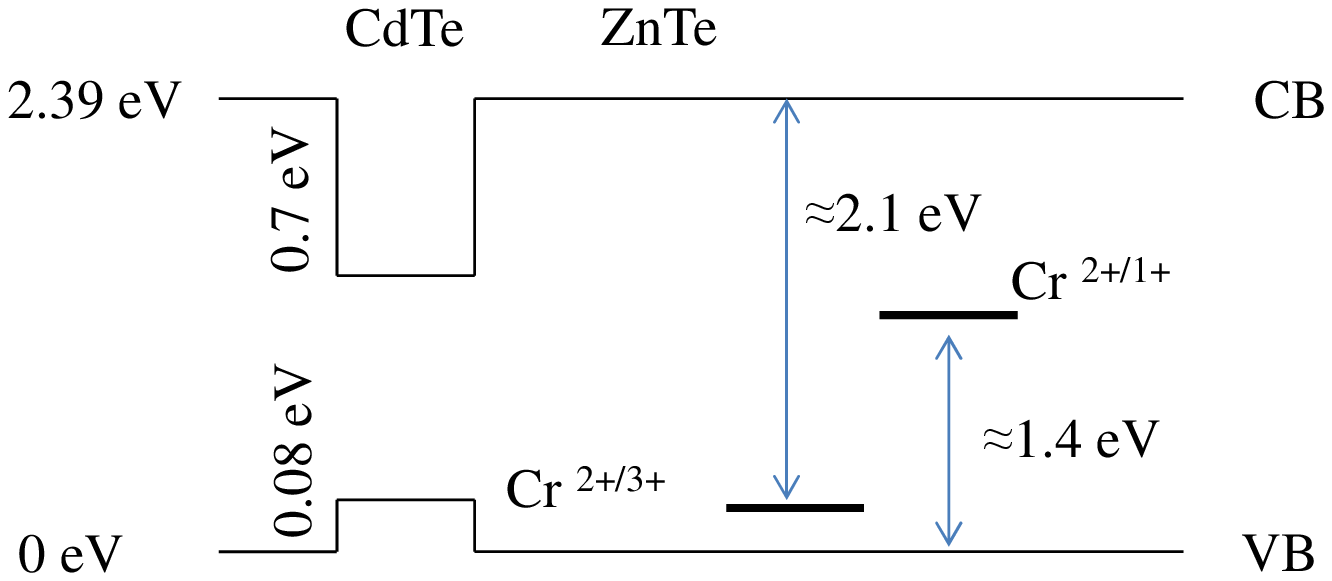}
\caption{Energy of the acceptor and donor levels of Cr in ZnTe and confinement energies in the CdTe QDs. Blue arrows are possible optical transitions.}
\label{Fig5}
\end{figure}

In an intrinsic ZnTe matrix, a Cr atom is preferentially incorporated as an isoelectronic Cr$^{2+}$ impurity where the 4\textit{s} outer electrons of
the atoms are shared in the crystal bond \cite{Kuroda2007}. However the donor level Cr$^{2+/3+}$  and the acceptor level Cr$^{2+/+}$ are both within
the ZnTe band gap (see Fig. \ref{Fig5}). According to near infrared spectroscopy measurements \cite{Dziesiaty1997}, the acceptor level is located
about 1.4 eV above the valence band. The position of the donor level has not been precisely determined but some spectroscopy data suggest that it is
situated about 2.1 eV below the conduction band, close to the top of the valence band \cite{Dziesiaty1997}. This value is not well known in bulk ZnTe
and could even be significantly modified for a Cr located in the vicinity of a CdTe QD in a region which is strained and likely form of a CdZnTe
alloy.

With this energy level configuration, the charge state of the Cr atom located in the ZnTe barrier can change by capturing a free electron or a free
hole coming from a chemical doping or injected optically \cite{Kuroda2007}. Under optical excitation below the ZnTe band gap, donor-type and
acceptor-type optical transitions can also lead to a change of the charge state of the Cr atom by promoting an electron from the valence band to the
atom or by transferring an electron from the atom to the conduction band.

Let us note that a fluctuation of the charge of the Cr between Cr$^{2+}$ and Cr$^{+}$ could also be possible for a Cr in CdTe as the Cr$^{2+/1+}$
acceptor level is within the CdTe band gap close to the conduction band \cite{Godlewski1980,Rzepka1993}. Such fluctuation between a 3$d^4$ (S=2) and
a 3$d^5$ (S=5/2) magnetic atom has never been observed in our experiments. If it occurs in a time scale faster than the PL integration time, the
exchange interaction of the fluctuating spin (between S=2 and S=5/2) with the confined carriers would result in a broad emission spectra for both
neutral and charged excitons. This broad spectra would be difficult to identify among all the broad emission lines observed in these Cr-doped samples
and coming from the interaction of the dot with several Cr atoms.

Cr$^{2+}$ is the neutral state of Cr in ZnTe and does not cproduce an electric field on a nearby QD. By capturing an electron, a Cr$^+$ ion is
created. Located close to a dot it attracts the confined hole and repels the electron. The opposite happens when the Cr capture a hole and a
Cr$^{3+}$ ion is created. A Cr atom located in the vicinity of a QD can then be seen as a punctual charge producing a fluctuating electric field
acting on the confined carriers.

In the CdTe/ZnTe QD system, the large band gap difference mainly contributes to a large conduction band offset and the electron is well confined. On
the other hand, the confinement of the hole is weak and for an exciton it is strongly influenced by the Coulomb potential created by the confined
electron. For a punctual charge located near the QD, the Coulomb interaction of the confined carriers with the localized charge is then expected to
be significantly different for the electron or the hole. This interaction can also deform the confined carriers wave function (especially for the
hole) and affects the electron-hole Coulomb interaction within the exciton. Both contributions will change the emission energy of the QD emission.

With a Cr atom located near the QD at a distance shorter than the lateral spatial extension of the confined carriers (a few nm) and with a charge
fluctuating between -e, 0 or +e, three different emission energies should be obtained for a neutral or a charged exciton, as observed in the reported
experiments. Though the optical transition always corresponds to the recombination of an electron-hole pair, the influence of the electric field
produced by a given localized charge could be different for the different excitonic complexes (exciton, biexciton, positively or negatively charged
exciton) because of the electric field induced modification of their binding energy \cite{Besombes2002}.

\section{Modelling of the influence of the fluctuating electric field of a localized charge on a QD}

To confirm that the fluctuations of the charge between 0, -e and +e of a single defect located close to a QD can be responsible for the observation
of three energy levels in the emission spectra, we use a variational model to estimate the energy shift induced by the localized charge. In this
simple model we neglect the possible presence of a permanent dipole for the confined exciton \cite{Finley2004}, a usually good approximation for flat
lens shape QDs.

The confinement potential for a CdTe/ZnTe QD is characterized by a significantly larger confinement for the electron than for the hole. In our model
the confinement is simply described by a finite quantum well along the growth axis and a truncated parabolic potential for the in-plane motion. A
detailed description of the potentials and trial wave-functions used in the variational model are presented in appendix A.

\begin{figure}[hbt]
\includegraphics[width=3.4in]{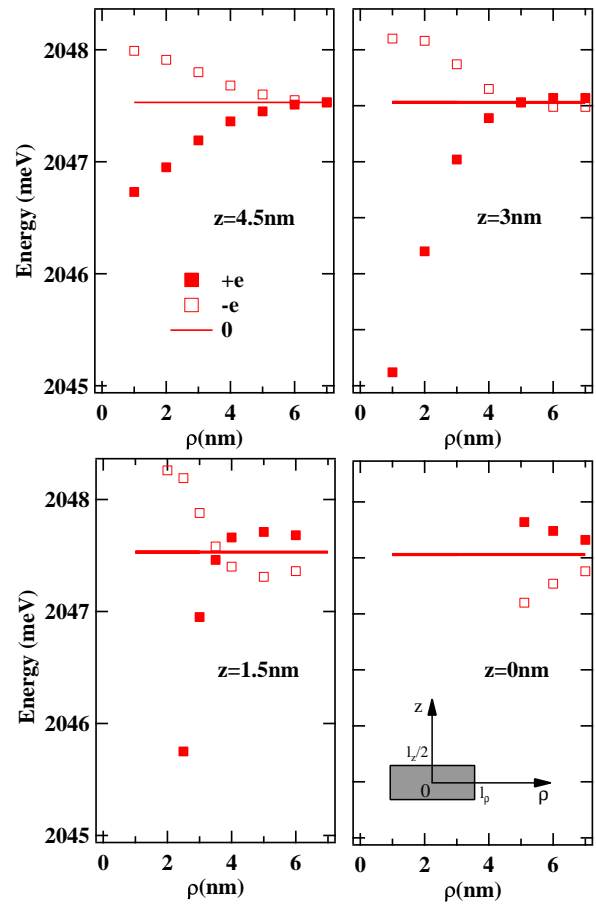}
\caption{(a) Calculated energy levels of the exciton as a function of the position of a charge +e or -e localized in the ZnTe barrier near the QD.
The evolution of the emission energy is presented as a function of the in plane distance $\rho$ for different fixed values of $z$ ($z$=0nm, $z$=1.5nm, $z$=3nm).
The distances $\rho$ and $z$ are measured from the center of the QD. The size of the QD is l$_z$=2.5 nm and 2l$_{\rho}$=10nm (see appendix A for a detailed description of the confinement potential). }
\label{Fig6}
\end{figure}

Results of the variational model for the transition energy of an exciton as a function of the position and value of a punctual charge are presented
in Fig.~\ref{Fig6}. This model first gives a good order of magnitude for the energy of the excitonic emission and above all shows that a charge $\pm
e$ located a few nm away from the QD can easily induce an energy shift of a few tens of $\mu$eV that can be detected in the optical spectra. For a
given charge state of the defect, the amplitude of the shift as well as its sign strongly depend on the position of the defect.

For a charge located above (or bellow) the QD close to its $z$ axis, the interaction is dominated by the Coulomb interaction with the strongly
localized electron and the energy is decreased (increased) for a charge +e (-e). The situation is reversed for a charge located in the plane of the
QD (z=0 nm in Fig.~\ref{Fig6}) where the energy shift induced by a charge +e is positive and negative for a charge -e.

This simple model shows that three energy levels separated by a few hundred $\mu$eV can be observed for the emission of an exciton in a QD located
near a single fluctuating charge. However, as already observed for a CdTe/ZnTe QD submitted to a random fluctuating electric field
\cite{Besombes2002}, the Stark shift induced by a given electric field can be significantly different for the different excitonic complexes (neutral
and charged). For instance, a weaker red shift was observed for the biexciton than for the exciton \cite{Besombes2002}. This different shift under
electric field results from a stronger reduction of the binding energy for the biexciton than for the exciton. A more pronounced reduction of the
binding energy can even be observed for the charged excitons leading to shift under electric field of different signs for the exciton and the charged
excitons.

This behaviour under electric field is consistent with the observation in our samples of a different splitting for the different excitonic complexes
in the QDs influenced by the fluctuating electric field of a single Cr atom. For instance a weaker splitting for the biexciton than for the exciton
is systematically observed (see for instance Fig.~\ref{Fig2} and Fig.~\ref{Fig3}), in agreement with the measurements reported in
\cite{Besombes2002}. This is an additional evidence permitting to attribute the complex emission structure observed in these Cr-doped QD samples to
the local fluctuation of an electric field.

\section{Influence of the optical excitation and external static electric field on the charge fluctuations of a Cr atom.}

\begin{figure}[hbt]
\includegraphics[width=3.25in]{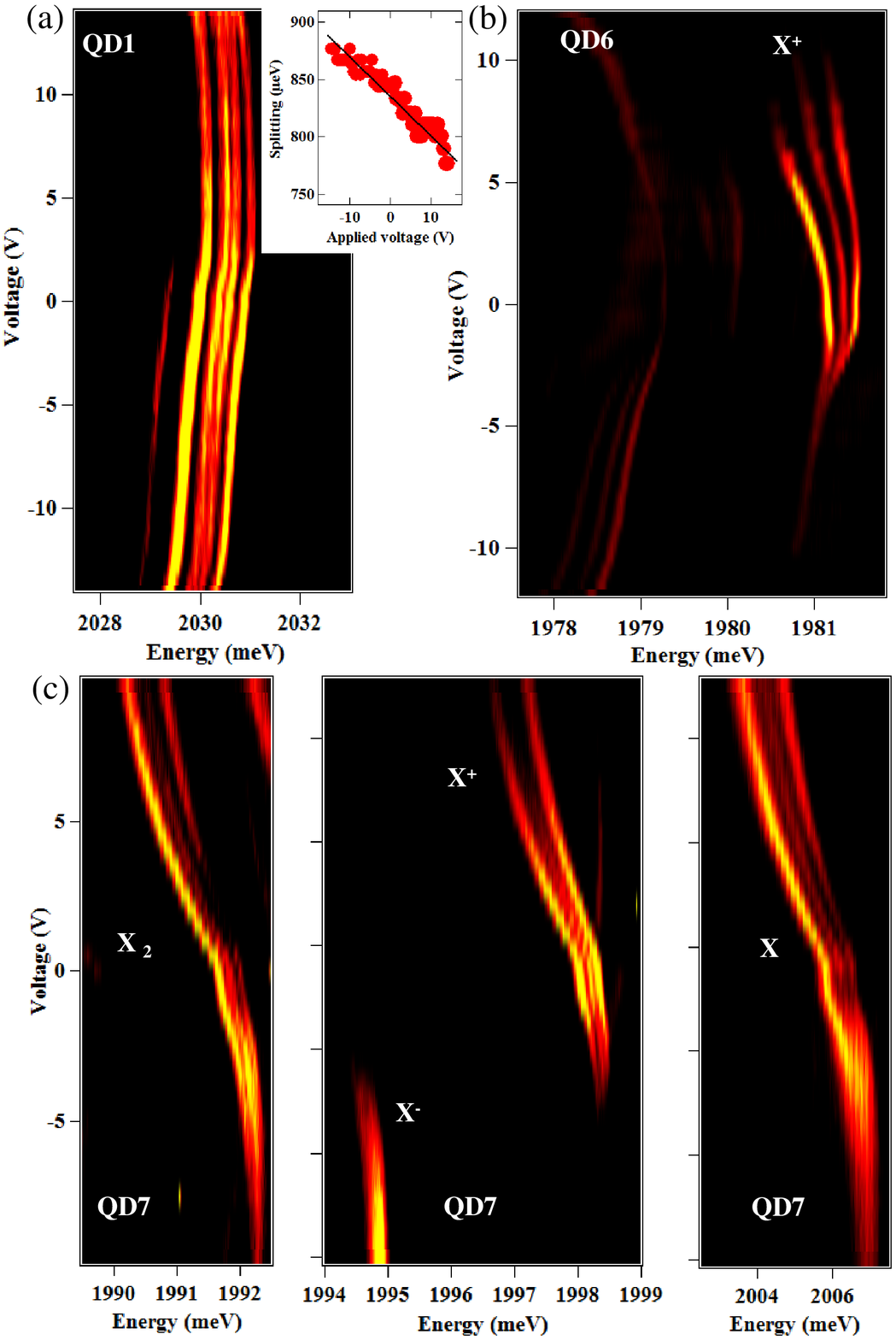}
\caption{(a) Applied bias voltage dependence of the PL of the exciton in the reference Cr-doped QD (QD1). The inset shows the voltage dependence of the exchange induced overall splitting of the exciton.
(b) Applied voltage dependence of the emission of the positively charged exciton in QD6.
(c) Applied voltage dependence of the emission of X, X$^+$ and X$_2$ in QD7.}
\label{Fig7}
\end{figure}

A static electric field applied along the growth axis of the QD can be used to modify the influence of the fluctuating charge of the Cr on the
confined carriers. The electric field dependence of the emission of a QD submitted to the charge fluctuation of a Cr atom are presented in
Fig.~\ref{Fig7} together with the electric field dependence of a reference magnetic QD containing a Cr atom (QD1 in Fig.\ref{Fig1}).

For a magnetic QD (Fig. \ref{Fig7}(a)), the applied uniform electric field produces a Stark shift of the exciton and a small modification of the
splitting between the low and high energy lines. The modification of the overall splitting of the emission results from a change of the exchange
interaction between the spin of the confined exciton and the spin of the Cr atom. The weak effect of the electric field on the splitting shows that
for such QD with a large carriers/Cr exchange interaction, the overlap between the atom and the confined carriers is not significantly affected by
the uniform electric field applied in the Schottky structure.

The effect of the static electric field is much more pronounced on the type of QDs submitted to the fluctuating electric field of a Cr located in the
barrier near the QD. This is particularly clear on X$^+$ where the splitting of the three lines under electric field can be increased at positive
bias or almost completely suppressed under a negative bias at the onset of the formation of the negatively charged exciton (Fig.~\ref{Fig7}(b)).

As presented for QD7 in Fig.~\ref{Fig7}(c), a similar change of the splitting can also be measured on the neutral excitonic complexes (X and X$_2$).
This effect is however usually more difficult to observe on the neutral species which are broadened by the electron-hole exchange interaction.

The application of an electric field through the Schottky gate mainly affects the weakly confined hole: it is attracted or repelled from the sample
surface depending on the bias voltage. The electric field induced variation of the splitting of the three lines is then likely to be due to a
significant modification of the interaction of the hole with the charge fluctuating Cr atom.

\begin{figure}[hbt]
\includegraphics[width=3.25in]{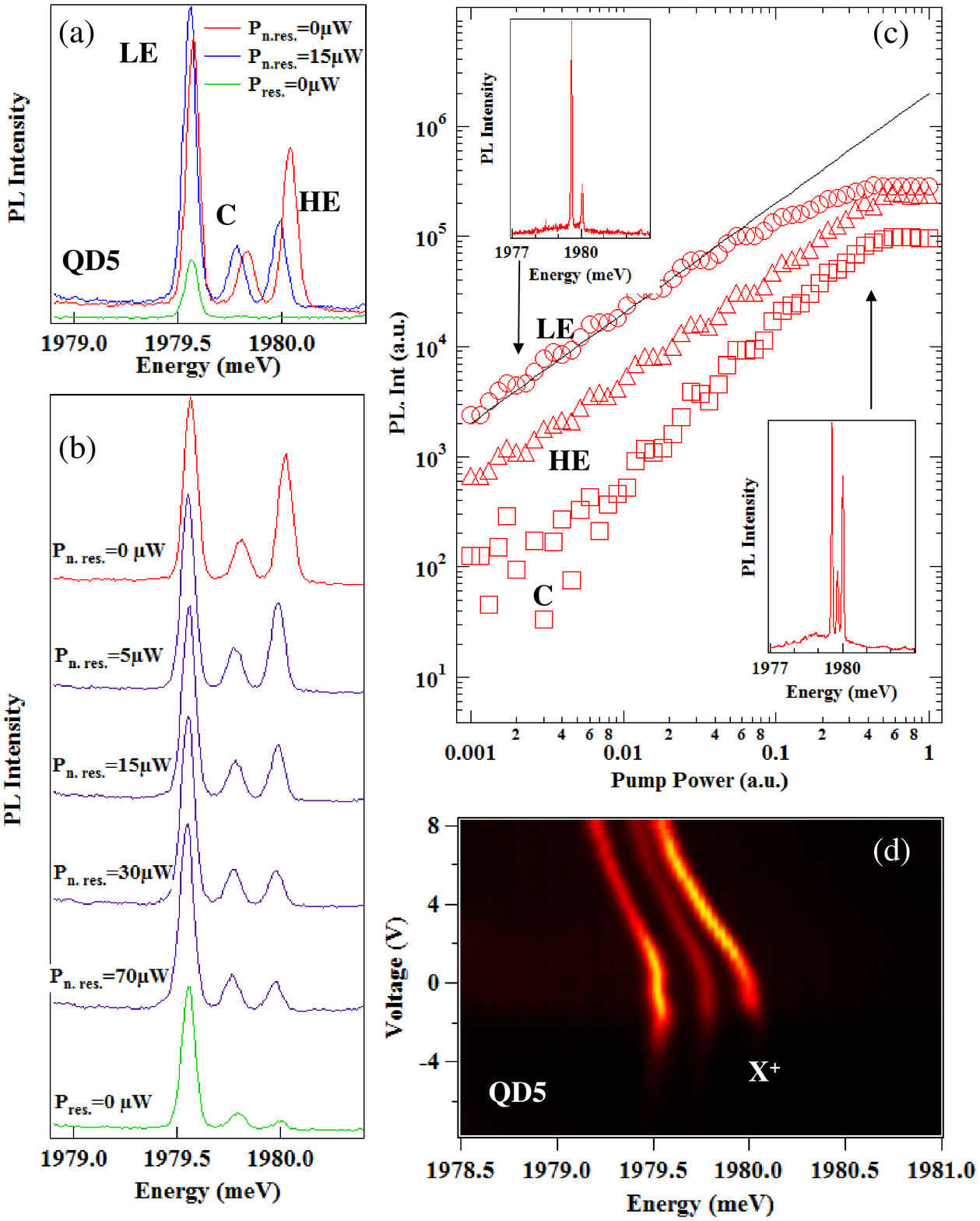}
\caption{(a) PL of the positively charged exciton in QD5 for a resonant (res.) excitation on an excited state alone (red), a weak excitation at 532 nm (n. res.) with $P_{n.res.}=15\mu W$ (green) and a combined resonant and non-resonant excitation (blue).
(b) Dependence of the intensity distribution of the positively charged exciton in QD5 for a fixed resonant excitation and variable non-resonant excitation $P_{n.res.}$.
(c) Intensity distribution on the three PL lines of the positively charged exciton (HE: high energy, LE: low energy, C: central) in QD5 as a function of the resonant excitation intensity. The black line is a linear fit.
(d) Intensity distribution in the PL of the charged exciton as a function of the applied bias voltage.}
\label{Fig8}
\end{figure}

The fluctuations of the charge of the Cr atom and its influence on the QD can also be modified by the wavelength of the optical excitation and its
intensity. This is illustrated in Fig.~\ref{Fig8} for the PL of X$^+$ in QD5. In this experiment, in addition to the excitation resonant with an
excited state of the QD (around 2076 meV) which gives the main contribution to the PL, a non-resonant excitation at 532 nm (2330 meV) close to the
band gap of ZnTe is added. Even if the contribution of the non-resonant excitation to the PL is very weak compared to the resonant PL, it
significantly changes the intensity distribution among the three lines of the charged exciton.

The most significant effect is a strong reduction of the intensity of the high energy line when the power of the non-resonant excitation is
increased. At high non-resonant excitation power the intensity of the central line is also slightly reduced. Under pure non-resonant excitation (532
nm in Fig.~\ref{Fig8}(b)), the PL mainly arises from the low energy line of the triplet of the charged exciton. Let us note that, as observed for the
application of a bias voltage, the presence of the non-resonant excitation reduces the splitting of the three lines (see Fig.~\ref{Fig8}(a)). The
non-resonant excitation creates carriers in the barriers which modifies the charge distribution in the Schottky structure and consequently the static
electric field already present at zero applied bias voltage.

It is reasonable to consider that the free carriers created in the ZnTe barriers by the non-resonant excitation at 532 nm tend to neutralize the
charged defects. Such change of the charge state by modulating the excitation wavelength was already observed for CdTe/ZnTe QDs
\cite{Leger2006,Leger2007}. The two-wavelength excitation experiments suggest then that the low energy line of the charged exciton likely corresponds
to the neutral configuration of the nearby Cr in ZnTe, Cr$^{2+}$, where no electric field are applied on the confined carriers. The two higher energy
lines would then correspond to the same charged exciton transition shifted by the local electric field of the Cr in either a Cr$^+$ or Cr$^{3+}$
configuration. This also suggests that the local electric field produced by a +e or -e charge induces a destabilisation (blue shift) of the
positively charged exciton. Such blue shift of the charged exciton was already observed on individual CdTe QDs under a random electric field produced
by charge fluctuations \cite{Besombes2002}. It was attributed to a significant reduction of the binding energy of the charged exciton induced by the
separation of charges.

The intensity distribution on the three lines of X$^+$ under resonant excitation also significantly depends on the excitation power. This is
illustrated on the excitation power dependence presented in Fig.~\ref{Fig8}(c). As for the non-resonant excitation, at low excitation intensity the
PL mainly arises from the low energy line suggesting that the charge fluctuations of the Cr are significantly reduced at low excitation intensity.

Resonant excitation on an excited state of the QD significantly below the band gap of the barriers can, in addition to the injection of carriers in
the QD, excite the Cr atom by promoting an electron from the valence band to create a Cr$^+$ or by transferring an electron from Cr$^{2+}$ to the
conduction band to create Cr$^{3+}$. Under excitation on an excited state of the QD (around 2076 meV for the experiments presented on
Fig.~\ref{Fig8}) the transition Cr$^{2+}$ to Cr$^+$ can be optically excited. This wavelength is at the edge of the Cr$^{2+}$ to Cr$^{3+}$ transition
which is only significantly excited at high excitation power. Even if we do not have definitive evidences, the Cr$^{3+}$ configuration likely
corresponds to the central line of the triplet of the charged exciton which completely disappears at low excitation intensity and presents a
super-linear excitation power dependence.

Let us note finally that the applied uniform electric field also slightly changes the charge fluctuation of the defect. This is observed as a change
in the relative intensities of the three lines of the positively charged exciton under electric field. This is presented for QD5 in
Fig.~\ref{Fig8}(d). Whereas at zero or negative bias the PL mainly arises from the low energy line of X$^+$, at positive bias the intensity is
transferred on the high energy line suggesting that under a static electric field a different charge state of the Cr is preferentially stabilized.

\section{Conclusion}

We have shown that in Cr-doped CdTe/ZnTe QDs samples, the emission of individual dots can be sensitive to the charge fluctuations of a single Cr atom
located nearby in the ZnTe barriers. For a single dot, three emission energies are then observed. They result from the energy shift induced by the
electric field of the three possible charge states of a Cr atom in ZnTe which fluctuate in a few hundreds of nanoseconds as revealed by
cross-correlation measurements. We have shown that the charge fluctuations of the Cr are affected by the intensity and wavelength of the optical
excitation. The influence of these charge fluctuations on a nearby QD can also be modulated by an applied static electric field. These results
suggest that during the growth process of these magnetic nano-structures, a significant amount of Cr is incorporated in the barriers in the vicinity
of the CdTe layer. This results from a diffusion of Cr atoms outside the CdTe layer or a segregation of the Cr atoms during the QD growth process.
The mechanism of Cr incorporation during the growth of these highly strained structure has to be studied in detail by additional structural
characterization.

\begin{acknowledgements}{}

This work was realized in the framework of the Commissariat \`{a} l'Energie Atomique et aux Energies Alternatives (Institut Nanosciences et
Cryog\'{e}nie) / Centre National de la Recherche Scientifique (Institut N\'{e}el) joint research team NanoPhysique et Semi-Conducteurs. The work was
supported by the French ANR project MechaSpin (ANR-17-CE24-0024) and CNRS PICS contract No 7463. V.T. acknowledges support from EU Marie Curie grant
No 754303.  The work in Tsukuba has been supported by the Grants-in-Aid for Scientific Research on Innovative Areas "Science of Hybrid Quantum
Systems" and for challenging Exploratory Research.

\end{acknowledgements}

\begin{appendix}

\section{Variational calculation of the influence of a single fluctuating charge on the energy of a confined exciton}

To estimate the influence of an individual localized charge on the energy of the exciton in a CdTe/ZnTe QD we use a variational method. The
confinement in our QDs is described by a potential that takes into account the difference of confinement regime in the plane and along the QD growth
axis. Along z, the potential is described by a quantum well of finite height $\Delta E_{c,v}$ and thickness $l_z$:

\begin{eqnarray}
V_{e,h}^z=\Delta E_{c,v}\Pi_{l_z}(z)
\end{eqnarray}

\noindent where $\Delta E_{c}=0.7eV$ and $\Delta E_{v}=0.08eV$ are the conduction band and valence band offset respectively and $\Pi_{l_z}$ the
rectangular function of full width $l_z$.

In the plane, the confinement is described by a parabolic and isotropic potential of lateral extension $2l_{\rho}$:

\noindent - for $|z|<l_z$/2

\begin{eqnarray}
V_{e,h}^{\rho}=\Delta E_{c,v}(\frac{\rho^2}{l_{\rho}^2}-1)\Pi_{2l_{\rho}}(\rho)
\end{eqnarray}

\noindent - for $|z|>l_z$/2
\begin{eqnarray}
V_{e, h}^{\rho}=0
\end{eqnarray}

In the variational calculation, the trial wave function of the electron $\psi_e(z,\rho)$ is decomposed into a Gaussian wave function in the QD plane

\begin{eqnarray}
\psi_e(\rho)= \frac{1}{\sqrt{\pi}\sigma_e}e^{-\frac{\rho^2}{2\sigma_e^2}}
\end{eqnarray}

\noindent where $\sigma_e$ is a variational parameter, and a wave function $\psi_e(z)$ corresponding to the ground state of a finite quantum well
along z. $\psi_e(z)=Be^{\kappa z}$ for $z < -l_z$/2, $\psi_e(z)=A \cos(k z)$ for $|z|<l_z$/2 and $\psi_e(z)=Be^{-\kappa z}$ for $z
> l_z$/2 with

\begin{eqnarray}
A=\sqrt{\frac{2}{l_z+2/\kappa}};B=\sqrt{\frac{2}{l_z+2/\kappa}}e^{\kappa l_z/2}k/k_0
\end{eqnarray}

\noindent and the wave vectors:

\begin{eqnarray}
\kappa=\sqrt{\frac{-2m_{e}^zE}{\hbar^2}};
k_0=\sqrt{\frac{2m_{e}^z\Delta E_c}{\hbar^2}};\\
k=\sqrt{\frac{2m_{e}^z(E+\Delta E_c)}{\hbar^2}}
\end{eqnarray}

\noindent with E the energy of the ground state in the quantum well (E $<$0 as the origin of energy is chosen at the bottom of the conduction band of
the barriers) controlled by the equation $\cos(k l_z/2)=k/k_0$.

As CdTe/ZnTe QDs present a weak valence band offset, the  confinement of the hole is strongly influenced by the coulomb attraction of the electron.
We chose for the hole a wave function decomposed on two Gaussians, one for the motion along z and one for the in-plane motion:

\begin{eqnarray}
\psi_h(z,\rho)=\frac{1}{\pi^{3/4}\sqrt{\sigma_z^h}\sigma_{\rho}^h}e^{-\frac{z^2}{2(\sigma_z^h)^2}}e^{-\frac{\rho^2}{2(\sigma_{\rho}^h)^2}}
\end{eqnarray}

\noindent where $\sigma_z^h$ and $\sigma_{\rho}^h$ are two variational parameters.

A single Cr atom with a charge $\pm e$ located at the position $\overrightarrow{r_{Cr}}(\rho_{Cr},\theta_{Cr},z_{Cr})$ induces a potential energy for
the confined carrier at position $\overrightarrow{r_i}$

\begin{eqnarray}
V_{Cr}(\overrightarrow{r_i})=\frac{\pm e^2}{4\pi\epsilon\epsilon_0|\overrightarrow{r_{Cr}}-\overrightarrow{r_i}|}
\end{eqnarray}

The total energy of the confined electron is then given by:

\begin{eqnarray}
H_e=\frac{\overrightarrow{p_e}^2}{2m_e^*}+V_e(\overrightarrow{r_e})+V_{Cr}(\overrightarrow{r_e})
\end{eqnarray}

In the variational calculation, we minimize the total energy of the electron to determine the variational parameter $\sigma_e$ and then take into
account the Coulomb attraction created by the confined electron $U_e(\overrightarrow{r_h})$ to determine the energy of the confined hole. The total
energy of the hole is then given by the Hamiltonian

\begin{eqnarray}
H_h=\frac{\overrightarrow{p_h}^2}{2m_h^*}+V_h(\overrightarrow{r_h})+U_e(\overrightarrow{r_h})+V_{Cr}(\overrightarrow{r_h})
\end{eqnarray}

\noindent with

\begin{eqnarray}
U_e(\overrightarrow{r_h})=e\int\frac{-e}{4\pi\epsilon\epsilon_0}\frac{|\psi_e(\overrightarrow{r_e})|^2}{|\overrightarrow{r_e}-\overrightarrow{r_h}|}\overrightarrow{dr_e}
\end{eqnarray}

This energy is minimized to find the level of the confined hole in the presence of the Coulomb potential of the electron.

The energy levels of the electron and the hole are then finally used to determine the energy of the excitonic transition in the presence of the
fluctuating localized charge +e, 0, -e.

\end{appendix}

\end{document}